\documentclass[manuscript]{aastex}  
\newcommand{\about}     {\hbox{$\sim$}}

\newcommand{\uM}                {\hbox{$u^*$}}
\newcommand{\gM}                {\hbox{$g^*$}}
\newcommand{\rM}                {\hbox{$r^*$}}
\newcommand{\iM}                {\hbox{$i^*$}}
\newcommand{\zM}                {\hbox{$z^*$}}
\newcommand{\ug}               {\hbox{$u^*-g^*$}}
\newcommand{\gr}               {\hbox{$g^*-r^*$}}
\newcommand{\ri}               {\hbox{$r^*-i^*$}}
\newcommand{\iz}               {\hbox{$i^*-z^*$}}
\newcommand{\urc}              {\hbox{$u^*-r^*$}}

\newcommand{\sd}{\ensuremath{\Box^{\circ}}}
\newcommand{\si}[1]{\ensuremath{_\textrm{\scriptsize{#1}}}}

\begin{document}
  

\title{  Color Separation of Galaxy Types in the Sloan Digital Sky Survey Imaging Data  }

\author{
Iskra Strateva\altaffilmark{1},
\v{Z}eljko Ivezi\'{c}\altaffilmark{1},
Gillian R. Knapp\altaffilmark{1},
Vijay K. Narayanan\altaffilmark{1},
Michael A. Strauss\altaffilmark{1},
James E. Gunn\altaffilmark{1},
Robert H. Lupton\altaffilmark{1},
David Schlegel\altaffilmark{1},
Neta A. Bahcall\altaffilmark{1},
Jon Brinkmann\altaffilmark{2},
Robert J. Brunner\altaffilmark{3},
Tam\'as Budav\'ari\altaffilmark{4,5},
Istv\'an Csabai\altaffilmark{4,5},
Francisco Javier Castander\altaffilmark{6},
Mamoru Doi\altaffilmark{7},
Masataka Fukugita\altaffilmark{8,9},
Zsuzsanna Gy{\H o}ry\altaffilmark{4,5},
Masaru Hamabe\altaffilmark{7},
Greg Hennessy\altaffilmark{10},
Takashi Ichikawa\altaffilmark{11},
Peter Z. Kunszt\altaffilmark{4},
Don Q. Lamb\altaffilmark{6},
Timothy A. McKay\altaffilmark{12},
Sadanori Okamura\altaffilmark{7},
Judith Racusin\altaffilmark{12},
Maki Sekiguchi\altaffilmark{8},
Donald P. Schneider\altaffilmark{13},
Kazuhiro Shimasaku\altaffilmark{7},
Donald York\altaffilmark{6}
}

\newcounter{address}
\altaffiltext{1}{Princeton University Observatory, Princeton, NJ 08544}
\altaffiltext{2}{Apache Point Observatory, P.O. Box 59, Sunspot, NM 88349-0059}
\altaffiltext{3}{Department of Astronomy, California Institute of Technology, 
Pasadena, CA 91125}
\altaffiltext{4}{Department of Physics and Astronomy, The Johns Hopkins 
University, 3701 San Martin Drive, Baltimore, MD 21218}
\altaffiltext{5}{Department of Physics, E\"{o}tv\"{o}s University, Budapest, 
Pf. 32, Hungary, H-1518}
\altaffiltext{6}{University of Chicago, Astronomy \& Astrophysics Center, 
5640 S. Ellis Ave., Chicago, IL 60637}
\altaffiltext{7}{Department of Astronomy and Research Center for the Early 
Universe, School of Science, University of Tokyo, Hongo, Bunkyo, Tokyo, 
113-0033 Japan}
\altaffiltext{8}{Institute for Cosmic Ray Research, University of Tokyo, 
Midori, Tanashi, Tokyo, 188-8502 Japan}
\altaffiltext{9}{Institute for Advanced Study, Olden Lane, Princeton, NJ 08540}
\altaffiltext{10}{U.S. Naval Observatory, 3450 Massachusetts Ave., NW, Washington, 
DC 20392-5420}
\altaffiltext{11}{Astronomical Institute, Tohoku University, Aoba, Sendai, 
980-8578 Japan}
\altaffiltext{12}{University of Michigan, Department of Physics, 500 East University, 
Ann Arbor, MI 48109}
\altaffiltext{13}{Department of Astronomy and Astrophysics, The Pennsylvania 
State University, University Park, PA 16802}

\begin{abstract}
 
We study the optical colors of 147,920 galaxies brighter than \gM  =
21, observed in five bands by the Sloan Digital Sky Survey (SDSS) over
\about 100 \sd\ of high Galactic latitude sky along the Celestial
Equator.  The distribution of galaxies in the \gr\ vs.\ \ug\
color--color diagram is strongly bimodal, with an optimal color
separator of \urc  =  2.22. We use visual morphology and spectral
classification of subsamples of 287 and 500 galaxies respectively, to
show that the two peaks correspond roughly to early (E, S0, Sa) and
late (Sb, Sc, Irr) type galaxies, as expected from their different
stellar populations. We also find that the colors of galaxies are
correlated with their radial  profiles, as measured by the
concentration index and by the likelihoods of exponential and de
Vaucouleurs' profile fits. While it is well known that late type
galaxies are bluer than early type galaxies, this is the first
detection of a local minimum in their color distribution.  In all SDSS
bands, the counts vs. apparent magnitude relations for  the two color
types are significantly different, and demonstrate  that the fraction
of blue galaxies increases towards the faint end.
\end{abstract}    


\section{Introduction}

It has been known at least since the late 1930s that colors of
galaxies reflect their dominant stellar populations  and thus
correlate with morphology  (Humason 1936,  Hubble 1936).  Morgan and
Mayall (1957) examined the spectra of 47 nearby galaxies and found
that stellar systems with spectra dominated by A, A+F, and F stars are
exclusively classified as Sc and Irr morphologically, F+G dominated
stellar systems  correspond to Sb galaxies, and the K stellar systems
are a mix of predominantly early type (E, S0, Sa) galaxies, with a
sizeable fraction of Sb spirals. De Vaucouleurs (1961) used a sample
of 148 galaxies to establish the dependence of galaxy color on
morphological  type. Since then, studies of the color distribution of
galaxies have helped reveal their dominant stellar  populations and
star formation histories. Typical studies of galaxy colors in recent
years have been based on samples of \about 1,000 galaxies. Fioc \&
Rocca-Volmerange (1999) used optical and near infrared colors of 1,000
galaxies to establish relations  between the colors, morphological
types, inclinations or shapes, and the intrinsic luminosities of
galaxies.  Ferreras {\em et al.}\ (1999) used color--magnitude and
color--color analysis of HST  photometry of \about 1,000 galaxies to
infer the existence of non-negligible star  formation in ellipticals
and bulges at  medium redshift (z \about 0.2). Brown {\em et al.}\
(2000)  studied the dependence of clustering of galaxies  on
color. They used a  catalog of \about $4\times10^5$ galaxies and
selection rules based on synthetic colors given by Fukugita {\em et
al.}\ (1995), and found that the galaxy correlation function is
strongly dependent on color, with red galaxies more strongly clustered
than blue galaxies by a factor of $\gtrsim$ 5 at small scales.

The Sloan Digital Sky Survey (hereafter SDSS, York {\em et al.}\ 2000)
is generating accurate photometry for an unprecedentedly large  and
uniform sample of galaxies, enabling us to expand galaxy color studies
and extend them in new directions.  The SDSS is a digital photometric
and spectroscopic survey which will  cover one quarter of the
Celestial Sphere toward the Northern Galactic cap, and produce a
smaller area ($\sim$225 \sd) but much deeper survey toward the
Southern Galactic cap. The photometric/astrometric mosaic  camera
(Gunn et al. 1998; see also Project Book \S 4\footnote {
http://www.astro.princeton.edu/PBOOK/welcome.htm}, ``The Photometric
Camera'') images the sky by scanning along great circles at the
sidereal rate. The flux densities of detected objects are measured
almost simultaneously in five bands ($u$, $g$, $r$, $i$, and $z$;
Fukugita {\em et al.}\ 1996) with effective wavelengths of 3543~\AA,
4770~\AA,  6231~\AA, 7625~\AA, and 9134~\AA \footnote{We refer to the
measured magnitudes  in this paper as $u^*, g^*, r^*, i^*,$ and $z^*$
because the absolute calibration  of the SDSS photometric system is
still uncertain at the $\sim 0.03^m$ level.  The SDSS filters
themselves are referred to as $u, g, r, i,$ and $z$.  All
magnitudes are given on the AB$_\nu$ system (Oke \& Gunn 1983). For
additional  discussion regarding the SDSS photometric system see
Fukugita {\em et al.}\ (1996) and Fan {\em et al.}\  (1999).}. The
telescope is also equipped with two double fiber-fed
spectrographs. Fiber plug plates  are individually drilled for each
field to accommodate 640 optical fibers of 3$''$ entrance diameter,
which feed the spectrographs.   The survey sky coverage of about $\pi$
steradians (10,000 \sd) will  result in photometric measurements of
\about 5$\times$10$^7$ galaxies, as well as \about $10^6$ moderate
resolution ($\lambda / \delta \lambda = 1800$) spectra  of galaxies
brighter than $r\si{petro} \approx 17.8$\footnote{For the definition
of Petrosian magnitude see Blanton {\em et al.}\ (2001), Yasuda {\em
et al.}\ (2001), and Strauss {\em et al.}\ (2001).},  covering the
wavelength range $3800-9200$~\AA. The morphological information from
the images currently allows robust star--galaxy separation to $\sim$
21.5$^m$ (Lupton {\em et al.}\ 2001, in preparation; Yasuda {\em et
al.}\ 2001).

The SDSS galaxy data have already been used in a number of studies.
Blanton {\em et al.}\ (2001) analyze 11,275 galaxies with redshifts
and photometry  to calculate the galaxy luminosity function (LF)  and
its dependence on galaxy properties such as surface brightness,
intrinsic color, and morphology. Yasuda {\em et al.}\ (2001) derive
the galaxy number counts, and Fischer {\em et al.}\ (2000) measure the
effect of galaxy--galaxy  weak lensing. A series of papers in
preparation (Zehavi {\em et al.}\ 2001,  Tegmark {\em et al.}\ 2001,
Dodelson {\em et al.}\ 2001, Szalay {\em et al.}\  2001, and Connoly
{\em et al.}\ 2001) analyze in detail the clustering of  galaxies in
SDSS and calculate the 3D power spectrum. Bernardi  {\em et al.}\
2001, study various scaling relationships in a sample of 9,000 early
type galaxies. In a nice complementary paper to the current work,
Shimasaku {\em et al.}\ (2001) investigate in detail the the colors,
effective size, and concentration parameter of SDSS galaxies based on
a sample of 456 bright objects classified visually into seven
morphological  types.  They perform important tests on consistency of
SDSS galaxy colors  with those obtained from both conventional BVRI
photometry and synthetic  colors calculated from template
spectroscopic energy distributions (SEDs)  of galaxies.

One of the scientific goals of the SDSS is to study the dependence of
galaxy  properties such as the luminosity function, size distribution,
evolution,  and large scale distribution on morphological type. While
morphological  types can be assigned with some certainty to nearby,
well resolved galaxies,  this is not possible for the fainter and more
distant galaxies imaged by SDSS. If we can find a relationship between
observed color and morphological type, this relation (if fairly
independent of K corrections) can in principle replace morphological
segregation in studies of galaxy properties and distribution all the
way to the limit at which we can do reliable star--galaxy separation.
In this paper, we study the color distribution of a large, uniform
sample of 147,920 galaxies detected in SDSS commissioning data, and
show that the \urc\ color distribution is bimodal, with a clear
separation between the  two classes down to the SDSS imaging faint
limit.  We describe the data samples  and analyze the color
distribution in Section 2, followed by a discussion of the
correlations  between color and conventional morphological and
spectral types in  Section 3. We present our conclusions in Section 4.

\section{          Data  Analysis              }

\subsection{           The Samples        }

In the analysis of the galaxy colors we use the model magnitudes as
measured by the SDSS photometric pipeline {\it Photo} (version 5.2,
Lupton {\em et al.}\ 2001). The model magnitudes are calculated by
fitting de Vaucouleurs and exponential models, convolved with the PSF,
to the two-dimensional images of galaxies in the $r$ band, and
computing the total magnitude corresponding to the better fit (see the
Appendix). This $r$ band fit is applied in all five bands, yielding
galaxy colors measured through the same aperture. The estimated
photometric errors are a function of magnitude: $\Delta m \approx
c_1+c_210^{0.2m}$, where the first term models the lower limit of the
error due to sky subtraction, and the second term models the photon
statistics. Those error estimates does not include the uncertainty in
the photometric calibration, which for this data is of the order of
$0.^m3$. The values of the $c_1$ and $c_2$ coefficients, obtained by a
linear fit for $c_2$ with $c_1$ set to the median value of the error
at the bright end ($m<16$), are given in Table 1. The photometric
errors are less than $0.^m1$ for $\uM \lesssim 19$, $\gM \lesssim 22$,
$\rM \lesssim 21$, $\iM \lesssim 21$, and $\zM \lesssim 18$. The
photometric errors at magnitudes of 16, 18, 20, 21, and 22 in all five
bands are given in Table 1. The quoted photometric errors are
consistent with those obtained by repeated observations.  We correct
the data for Galactic extinction determined from the maps given by
Schlegel, Finkbeiner \& Davis (1998). Typical values for the
high-latitude regions discussed in this work are \hbox{$A_{r^*}$ =
0.$^m$05 -- 0.$^m$15 ($A_{r^*} = 0.84 A_V$)}.

We discuss three galaxy samples, termed {\it photometric}, {\it
spectroscopic},  and {\it morphological}.  The {\it photometric}
sample is a magnitude limited (\gM $\leqslant$ 21) sample of  147,920
galaxies over 101.4 \sd\ of SDSS imaging data (equatorial  run 756)
obtained on 1999, March 22. The sample includes galaxies from five  of
the six non-adjacent 13.5$'$ wide strips along the Celestial Equator
with $-1.2687^\circ <$ $\delta\si{J2000}$ $<1.2676^\circ$ and 9$^h$
40$^m$  $<$ $\alpha\si{J2000}$ $<$ 15$^h$ 42$^m$. The photometric
errors in the \gM  $\leqslant$ 21 sample are typically less than
0.$^m$03 in \gM, \rM, \iM,  0.$^m$06 in \zM, and increase to \about
0.$^m$2 in \uM. This sample is used  to study the distribution of
galaxies in SDSS color space.

The {\it spectroscopic} sample, used for the detailed comparison of
color and spectral classification, contains 500 galaxies from a single
SDSS spectroscopic plate (plate number 267, obtained on February 22,
2000, consisting of 4$\times$15 min exposures). Of the 500 galaxies,
443 have $r^*\si{petro} < 17.8$ and are part of the main galaxy
spectroscopic sample (for more details on the galaxy spectroscopy
target selection, see Strauss \emph{et al.}\ 2001, in preparation). An
additional 57 galaxies are part of the Luminous Red Galaxies (LRGs)
sample, comprised of fainter, $0.25 < z < 0.50$, ellipticals
(Eisenstein \emph{et al.}\ 2001, AJ submitted).  The galaxies in the
spectroscopic sample are distributed in a circle with radius
1.49$^{\circ}$, centered on $\alpha\si{J2000}$ = 9$^h$ 50$^m$,
$\delta\si{J2000}$ = 0$^\circ$.  The photometric data for these
galaxies were obtained in SDSS commissioning runs 752 (obtained on
March 21, 1999) and 756.  The seeing FWHM in both runs was variable
between 1 and 2$''$, with the median value around 1.5$''$.

The {\it morphological} sample is a subsample of 287 bright galaxies
(\gM $<$ 16)  from the photometric sample, which we have classified by
eye,  and which allows studies of the correlation between color and
visual  morphology.

\subsection{         The Galaxy Color Distribution    }

\begin{figure}
\plotone{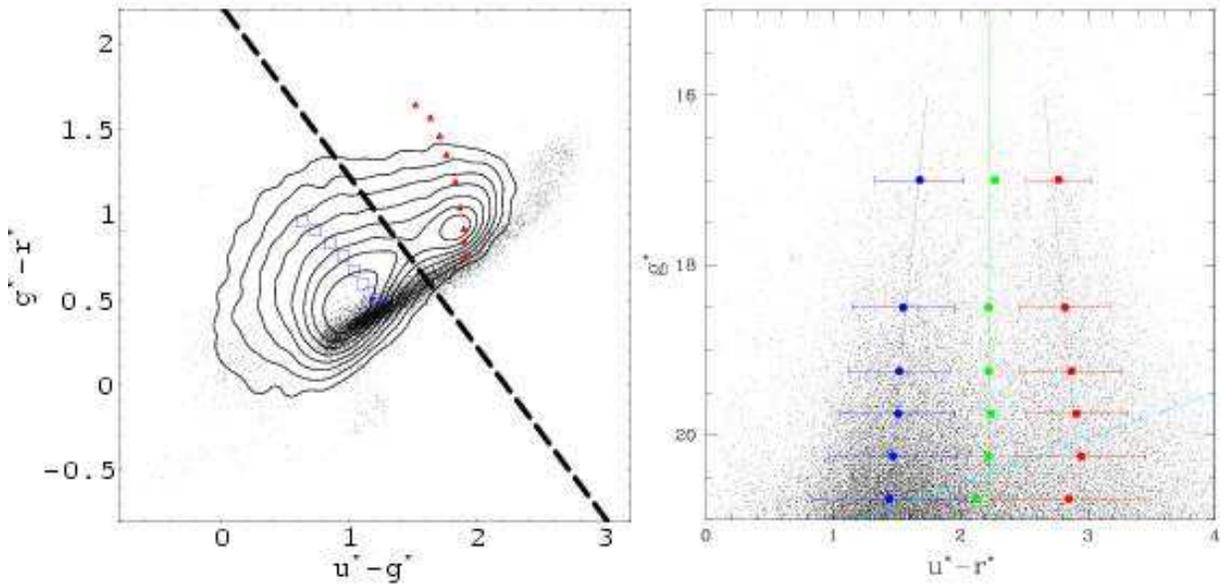}
\caption{ Left panel: Distribution of galaxies (contours) and stars
(dots) in the \ug\ vs.\ \gr\ color--color diagram. The contours
enclose  $\sigma$/4 (20.8\%) to $2\sigma$ (95.5\%) of all galaxies, in
steps of  $\sigma$/4 ($\sigma$ corresponds to the equivalent Gaussian
distribution). The thick long-dashed line is the \urc  =  2.22
separator.  The evolution of spiral (blue open squares) and elliptical
(red filled triangles) theoretical colors are given for $0 < z < 0.4$ at
every 0.05 in redshift.  Right panel: \urc\ vs.\ \gM\ color--magnitude
diagram of the photometric sample. Solid circles show positions of the
red and blue peaks and the  separator at each mean \gM\ of six
subsamples (see text). Thick lines give  linear regressions to each
peaks' variation, while the green short-dashed vertical line is \urc  =
2.22 separator. The slanted cyan long-dashed line is a \uM  =  22 cut.
\label{CmdCcd}} 
\end{figure}

The color--magnitude and color--color diagrams of galaxies in the
photometric  sample are presented in Figure \ref{CmdCcd}. The left
panel displays the \gr\ vs.\  \ug\ color--color diagram. The
distribution of galaxies from the photometric  sample is shown as
contours. For a detailed comparison between galaxy colors in the  SDSS
and various other photometric systems see Fukugita \emph{et al.} 1995;
for mean galaxy colors in the Cousins VRI photometric system as a
function of  Hubble T--stage, see Buta \& Williams, 1994. A sample of
stars with 15 $<$ \uM $<$ 18 extracted  from the same area of the sky
as the galaxy photometric sample is plotted as dots for
comparison. The \uM\ magnitude limits on the stellar sample were
selected to ensure a high signal-to-noise ratio (for more details
about the color distribution  of stars see, \emph{e.g.}, Fan 1999;
Finlator {\em et al.}\ 2000 and references therein).  The galaxy
distribution has two peaks, with the line connecting them almost
perpendicular to lines of constant \urc. This suggests that the \urc\
color is nearly optimal for separating galaxies into the two color
types.

The \gM\ vs.\ \urc\ color--magnitude diagram for the photometric
sample of galaxies is presented in the right panel of Figure
\ref{CmdCcd}. We will refer to the subsample of galaxies on the left
of the green short-dashed line as ``blue''  and the one on the right as
``red''. When plotted as a histogram, the \urc\ color distribution has
two maxima separated by a well-defined minimum. The positions of these
three extrema are only weakly dependent on the  sample magnitude
cut. We quantify the dependence of the three extrema  on the sample
magnitude limit by binning the photometric sample in \gM\ and  fitting
a sum of skewed ``Lorentzian'' profiles\footnote{Each ``Lorentzian''
is proportional to $[1+(x-x_o)^2/(ax+b)^2]^{-1}$. The exact form of
the fitted function is not critical, as long as it models the data
accurately.} to the \urc\ color distribution of the resulting
subsamples.  The blue and red peaks are then given by the maxima of
the ``Lorentzians''. We define the separator between the two as the
point at which the two ``Lorentzians'' (with areas individually
normalized to unity) have the same value.  Note that this is  not
equivalent to finding the minimum between the blue and red peaks,
since the ratio of red to blue galaxies is a function of  sample
magnitude cut, and decreases for fainter limiting magnitudes.  Sample
selection using $g$ band limiting magnitudes or redder bands
guarantees a sizeable fraction of red galaxies in fainter samples.
The \uM = 22 cut, represented by the slanted cyan long-dashed line in  the
right panel of Figure \ref{CmdCcd}, strongly decreases the number of
red relative to blue galaxies at fainter magnitudes.

The fitted positions of the three extrema for six magnitude subsamples
($16 < \gM < 21$) are plotted as filled circles in the right panel of
Figure \ref{CmdCcd} for each mean value of the \gM\ bin. The error
bars indicate the FWHMs of the best-fit ``Lorentzians''. The straight
lines fitted through those points are $(\urc)\si{blue} = 2.72 -
0.062\,\gM$ for the blue peak, and $(\urc)\si{red} = 2.17 +
0.035\,\gM$ for the red peak.  The variation in the position of the
\urc\ separator is very small (0.$^m$1) over the \gM\  range, showing
that this criterion for separating blue and red galaxies is valid over
a large range of magnitudes. The green vertical line through the separator
points is \urc = 2.22, corresponding to the mean and median of the six
fits.

Two trends in the \urc\ color distribution with fainter \gM\
magnitudes are visible in the right panel of Figure \ref{CmdCcd}: 1.\
a shift of the blue peak towards bluer and of the red peak towards
redder \urc\ colors, quantified by the line fits given above, and 2.\
an overall increase of the density of blue galaxies relative to the
red. The variation of the color distribution of the blue and red
subsamples with apparent magnitude is due to the fact that we sample
galaxies at increasing redshifts when selecting fainter magnitude
cuts.  The variations are caused by the color variation with
increasing redshift (K-corrections and galaxy evolution) and the
dependence of the sample on the galaxy number counts and luminosity
function.  Qualitatively, based on K-corrections alone (Fukugita {\em
et al.}\ 1995), we expect elliptical galaxies to get redder with
increasing redshift ($z \lesssim 0.3$) which is consistent with the
observed slope of the $(\urc)\si{red}$ line in the right panel of
Figure \ref{CmdCcd}.  Based on K-corrections only, we would expect the
blue galaxies to get initially redder in \urc\ by a few tenths of a
magnitude and then stay fairly constant with increasing redshift for
$z \lesssim 0.5$\footnote{Here we already anticipate the
identification  of the blue peak with spiral stellar population and
the red peak with  elliptical; see bellow.}. If we include galaxy
evolution through stellar population synthesis, we expect the \urc\
color of blue galaxies to stay almost constant up to z \about 0.4.
The color--color diagram of Figure 1 presents the expected \ug\ vs.\
\gr\ color evolution for a late spiral (blue open squares) and an
elliptical galaxy (red filled triangles). The symbols are plotted in 0.05
redshift intervals from z = 0 (redder \ug\ and bluer \gr) to z = 0.4
(bluer \ug\ and redder \gr). Beyond z \about 0.4, galaxies in the SDSS
photometric system evolve roughly perpendicular to the \urc = constant
cut in a \ug\ vs. \gr\ color--color diagram, and the \urc\ color is
not  a good separator. These evolutionary tracks were computed using
the evolutionary synthesis model PEGASE (Fioc \& Rocca-Volmerange
1997). The evolutionary prescriptions (star formation rate, initial
mass function and metallicity) were selected to reproduce the spectra
of nearby galaxies at z = 0.  The color evolution with redshift is
almost parallel to \urc = 2.22 (thick dashed line), keeping the
separator fairly constant with redshift up to z \about 0.4.  However,
the observed \urc\ color of the blue peak gets bluer with fainter
magnitudes (see the right hand side of Figure 1), departing from the
expected evolutionary behavior of a late spiral galaxy. This departure
is most likely due to an increasing degree of star forming activity
(for which the \urc\  color is a sensitive index) with redshift,
\emph{i.e.}, a larger number of galaxies with higher star formation
activity than the modeled galaxy presented in the Figure are seen at
high redshift than at low redshift.

\begin{figure}
\plotone{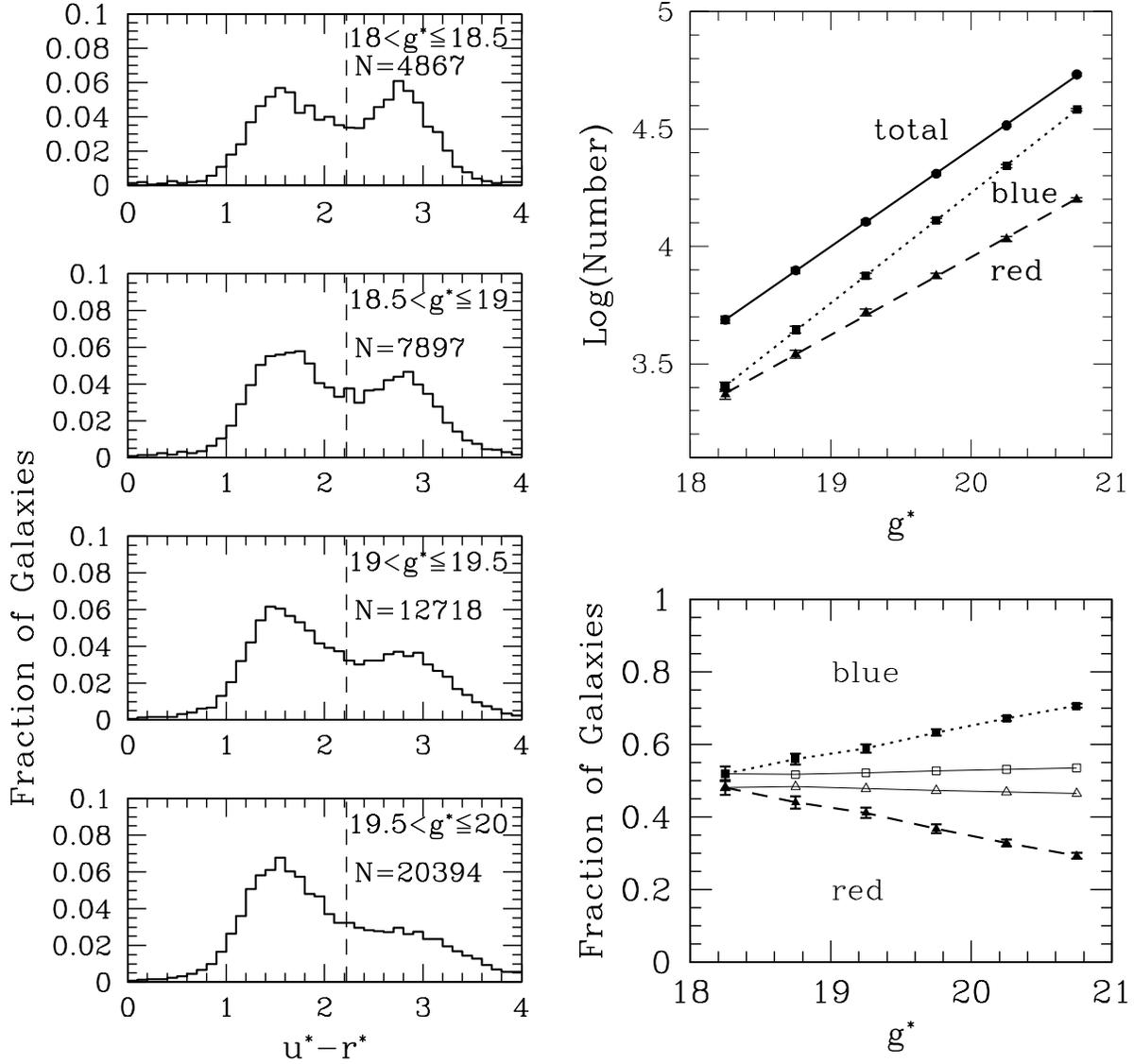}
\caption{Left panel: \urc\ color distribution as a function of \gM\
magnitude of the galaxy sample. Top right panel: The fraction of blue
galaxies (filled squares) increases relative to the red (filled
triangles) for fainter  \gM\ samples. Bottom right panel: photometric
errors cannot account for the dependence of the red and blue galaxy
fractions on magnitude cut. The open symbols correspond to the
predicted fraction (assuming only photometric errors  change with
magnitude), the filled symbols to the observed.\label{hist}}
\end{figure}

The second trend of the \urc\ color distribution with magnitude,
namely, the increase of the fraction of blue galaxies, is displayed in
the four color histograms for different \gM\ magnitude bins in the
left panel of Figure \ref{hist}. For $18 < \gM < 21$, the slopes of
the blue and red galaxy number counts in the top right panel are 0.47
and 0.33 respectively, amounting to a factor of \about 2 increase in
the number ratio of blue to red galaxies. Since K correction acts in
the opposite direction (\emph{i.e.}, to increase the number of red
galaxies), it cannot be responsible for the observed increase of blue
galaxies.  Moreover, this trend is present if we bin the data in \rM,
\iM,  or \zM\ bands, which sample the redder stellar populations.  Two
instrumental effects could cause the increase of the blue fraction as
measured in apparent \gM\ magnitude bins: the galaxy color
distribution could get wider for fainter red galaxies due to increased
photometric errors in the \uM\ band (which would affect red galaxies
more strongly, since they are fainter in \uM), or alternatively red
galaxies could be artificially ``leaking'' towards blue \urc\ colors,
because the asinh magnitudes (Lupton {\emph et al.}\ 1999) used by
SDSS cannot get fainter than a limit determined by the sky brightness
in that band.  We investigated both possibilities by assuming no
evolution and no K correction, \emph{i.e.}, the intrinsic distribution
of galaxies in faint magnitude samples is similar to the observed at
the bright end ($18.0<\gM<18.5$), and allowing the galaxies  to spread
as random Gaussian deviates in the flux with fainter apparent
magnitude.  Using the estimates for the photometric errors in \uM\ and
\rM\ as a function of magnitude given in Table \ref{tab1}, and $\uM(0)
= 24.63$ as the zero-point of the flux in the asinh magnitude
definition  (see Stoughton {\em et al.}\ 2001, in preparation), we
simulated the changes  in the \urc\ color distribution with \gM\
magnitude due to those two effects.  The lower right panel of Figure
\ref{hist} compares the simulated and observed fractions of red and
blue galaxies, demonstrating that those effects are not sufficient to
explain the change in the observed distribution. We found that the
increased photometric errors cannot explain the large difference
observed between the two peaks. The $\uM(0) = 24.63$ sky limit has a
negligible effect for the galaxies in our photometric sample, since
even the reddest are more than a magnitude brighter than this
limit. Increasing the photometric errors  by 50\% and/or lowering the
value of the sky limit to $\uM(0) = 24.00$  do not help in
reconstructing the observed change in the color distribution  with
apparent magnitude, suggesting that we are indeed seeing an increased
number of blue galaxies with fainter apparent magnitude. We conclude
that the observed evolution of the \urc\ color distribution with
apparent magnitude is a real evolutionary effect, caused by the
dependence  of the red and blue galaxy luminosity functions or
comoving volume number density on  redshift. The detailed modeling
needed  to establish this result quantitatively is beyond the scope of
this work.

\begin{figure}
\plotone{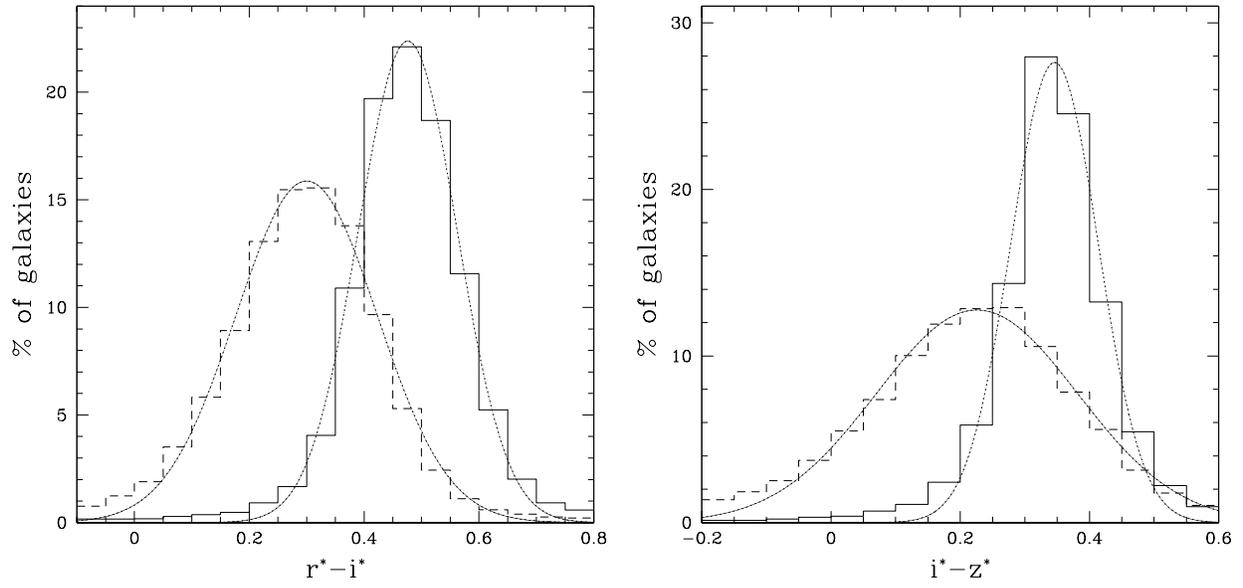}
\caption{\hspace{5mm}\ri\ and \iz\ color distributions for two
subsamples  separated by their \urc\ color (see text). The smooth
curves are  Gaussian fits to the data.\label{RiIzHist}}
\end{figure}

The two panels in Figure \ref{RiIzHist} compare the \ri\ and \iz\
color distributions  for the two \urc\ color-selected subsamples with
\gM $<$ 21. The histograms  represent the data distributions and the
curves show Gaussian fits.  It is evident that these two colors are
quite similar for ``blue'' and  ``red'' galaxies, with significant
overlap between the two subsamples.  The peak separation in both
colors is only about 0$^m$.1$-$0.$^m$15,  much smaller than the wide
\urc\ color peak separation ($\ga$ 1$^m$).  Since the \rM, \iM, and
\zM\ fluxes of galaxies are dominated by the old/low-mass stellar
populations present in all morphological types, it is indeed expected
that the \ri\ and \iz\ colors will not show much difference  for early
and late type galaxies.

\begin{figure}
\plotone{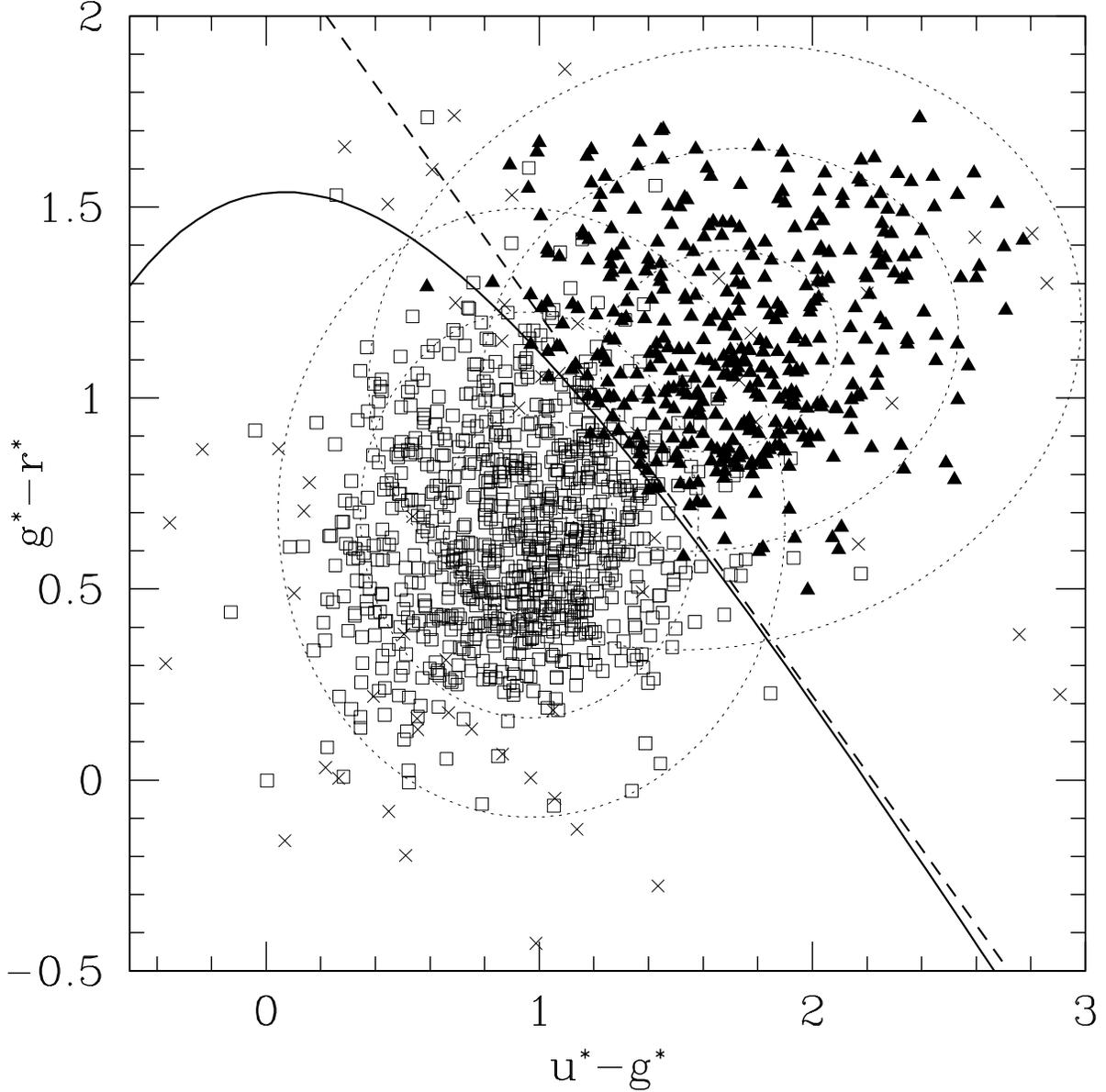}
\caption{Comparison of the Bayesian cut and the $u^*-r^*$ cut in a
projection of the  four-dimensional color space. Open squares and
solid triangles represent the two  main classes found by the
clustering algorithm (sparse sample). The crosses belong to the
outlier class. The thin dotted ellipses are the 1, 2 and 3$\sigma$
contours  of the projected class probability ellipsoids. The diagonal
dashed line is the $u^*-r^* = 2.22$ plane which closely follows the
Bayesian separator (solid line).\label{bayesCut}}
\end{figure}

However, the fact that the \ri\ and \iz\ colors are not identical for
the two \urc\ color selected galaxy types suggests that it may be
possible to use this additional information to perform a better galaxy
classification based on all four SDSS colors.  We used the program
AutoClass (Goebel {\em et al.}\ 1989; Cheeseman \& Stutz 1996) for an
unsupervised search for structure in the galaxy color
distribution. AutoClass employs Bayesian probability analysis to
automatically separate a given database into classes, and is an
efficient tool for analyzing multidimensional color diagrams. For
example, Ivezi\'c \& Elitzur (2000) used AutoClass to demonstrate that
the sources from the IRAS Point Source Catalog belong to four distinct
classes that occupy separate regions in the four-dimensional space
spanned by IRAS fluxes.  We searched for self-similar classes in the
galaxy color distribution by using a random subset of 25,000 galaxies
from the photometric sample. While the algorithm proposed 4 distinct
groups, most of the galaxies (82\%) are included in only two classes.
One of the remaining two classes represents outliers (5\%), and the
fourth one shows considerable overlap with one of the first two
classes.  We conclude that the bimodality is an excellent description
of the galaxy distribution in the SDSS color space even when all four
colors are used.  The boundary between the two galaxy types inferred
from the \urc\ distribution diagram is strongly supported by the
AutoClass results. Figure ~\ref{bayesCut} compares the Bayesian cut
(solid line) and the \urc\ cut (dashed line). The close agreement
between the unsupervised classification and the simple \urc\ color cut
is evident.

\section{Colors and Morphology}

\subsection{ Spectroscopic and Visual Classification }

The data presented in the previous section indicate that the \urc\
color  distribution is bimodal, and that galaxies can be divided into
``blue'' and ``red''  subsamples, as expected based on the differences
in the dominant stellar  populations for different morphological
galaxy types (\emph{e.g.}, de Vaucouleurs 1961).  In this section we
use independent morphological classification schemes to show  that the
blue galaxies are indeed dominated by late types (spirals) while the
red galaxies are dominated by early types (ellipticals).  This is
achieved by classifying a subsample of 287 galaxies using visual
appearance at the bright end ($\gM < 16$) and a fainter
($\rM\si{petro} < 17.7$)  subsample of 500 galaxies using spectra, and
comparing the results to the  separation based on \urc\ color.

\subsubsection{ Spectroscopic Classification and the \urc\ Color Distribution  }

\begin{figure}
\plotone{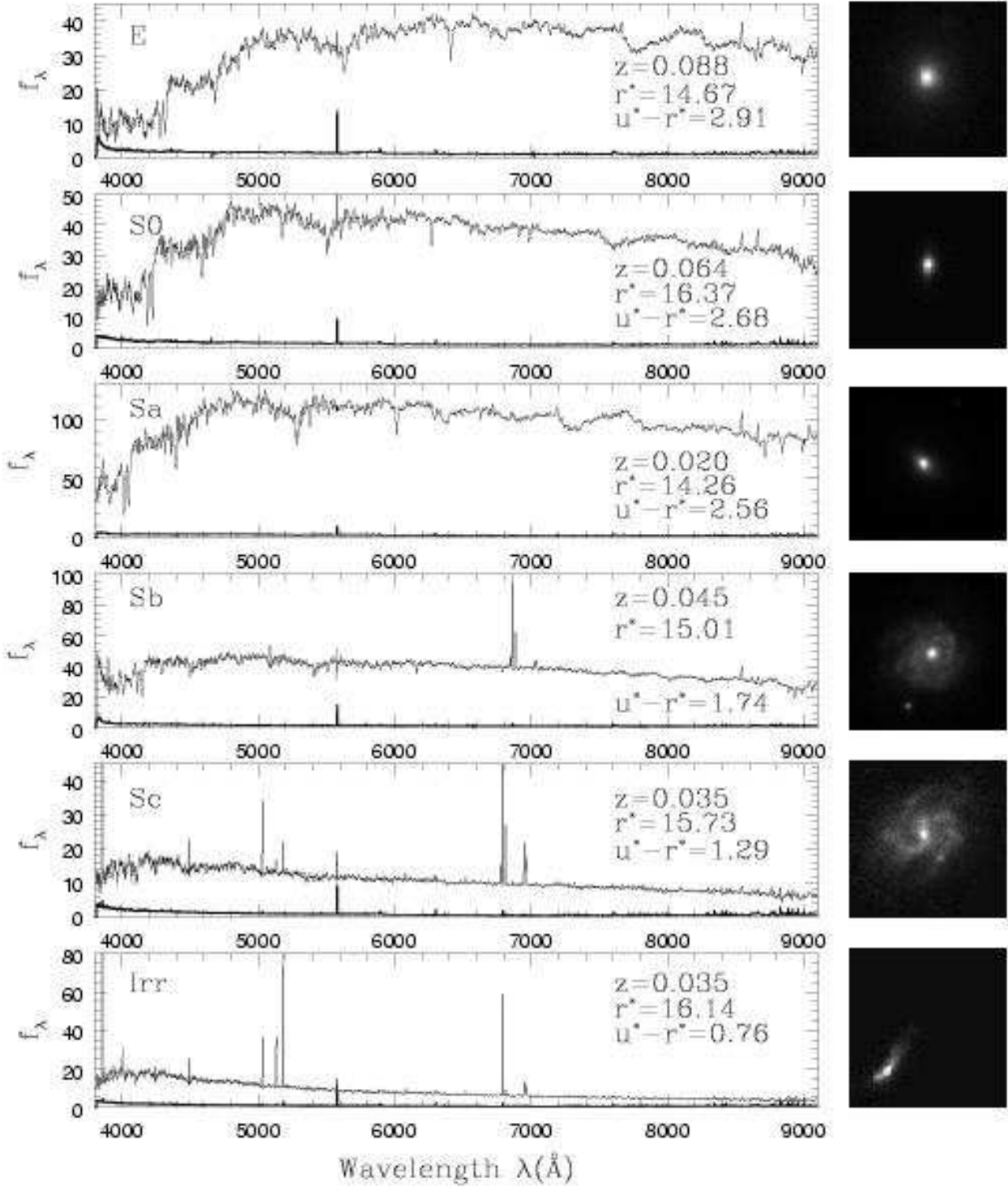}
\caption{Six galaxies from the spectroscopic sample representative of
the  different classes and their corresponding \gM\ band images. The
spectra are smoothed over 5\AA\ and the lower curve represents the
noise in each spectrum. Each image  is 40 $\Box''$.\label{spvis}}
\end{figure}

The 500 galaxies in the spectroscopic sample were classified by
visually comparing their spectra with templates from Kennicutt's
spectrophotometric atlas\footnote{Another approach to spectral
classification is to search for structure in the distribution of
various parameters, \emph{e.g.}, the expansion coefficients from the
Principal Component Analysis. This is beyond the scope of this
communication, and will be attempted in a separate publication.}
\cite{k1}. This classification is based on the relative strengths of
the H and K CaII absorption lines (3934~\AA, 3969~\AA), and the
H$\alpha$ (6563~\AA), [NII] (6583~\AA), [OII] doublet (3727~\AA),
H$\beta$ (4861~\AA), and [OIII] doublet (4959~\AA, 5007~\AA) emission
lines, when present. The spectral classification did not make use of
the equivalent widths (EWs) or the fluxes of the emission lines in a
quantitative way.  The galaxies are separated into six types: E(0),
S0(1), Sa(2), Sb(3), Sc(4), and Irr(5). Examples of the SDSS spectra
for the six different classes and \gM\  band images for those are
presented in Figure \ref{spvis}. In order to estimate the accuracy of
the classification, two of us (Strateva and Strauss) classified the
galaxies independently, and agreed in 410/500 cases to $\pm 1$ class.

The correspondence between spectral classification and \urc\ color is
shown in the left panel of Figure \ref{Cmds}. For simplicity, the
galaxies are grouped into ``early'' types (E, S0, Sa), shown as filled
triangles, and ``late'' types (Sb, Sc, and Irr), shown as open
squares.  Histograms of the \urc\  distribution for the spectroscopic
sample broken into four subclasses (E/SO, Sa, Sb, and Sc/Irr) are
given in the left panel of Figure \ref{ColHist}.  Practically all
(97.6\%) galaxies spectroscopically classified as early types have
\urc $\geqslant$ 2.22, and the remaining 3.1\% are early spirals (Sa)
with \urc $>$ 2.05. That is, we find no examples of spectroscopically
classified early type galaxies bluer than \urc = 2.05.  The galaxies
classified spectroscopically as late type show more scatter in their
colors.  While 153 of the 210 late type galaxies (73\%) have \urc $<$
2.22, and over 90\% (190/210) have \urc\ color bluer than 2.5, there
are still a small number (20/210) of the spectroscopically late
galaxies with 2.5 $<$ \urc $<$ 3.0 and presumably low star formation
rate and/or internal reddening (the images for  the majority of these
are consistent with either face--on spirals or ellipticals).

Kochanek {\em et al.}\ (2000) caution that spectroscopic
classification based on fixed small apertures will systematically
misclassify large angular diameter, late type galaxies as early types,
since their spectra sample mainly the bulge. Indeed 475 out of the 500
galaxies in our spectroscopic sample have more than 50\% of their
light outside the 3$''$ SDSS diameter fibers. We inspected visually
the 121 galaxies among those 475 that have early type spectra, and
found that 29 show a visual class later than their spectral class
(15/29 visually Sa/Sb galaxies show E/SO spectra, and 14/29 SO/Sa
galaxies show E/SO spectra), suggesting that the miss-classification
due to this  effect is about 6\%.

\subsubsection{ Visual Classification and the \urc\ Color Distribution }

The visual classification is based on the optical appearance in the
\gM\ band, which is the closest of the SDSS bands to the standard B
band used for classification. We classified the 287 bright galaxies
($\gM < 16$) from the morphological sample into the same six types as
in the spectroscopic sample: E(0), S0(1), Sa(2), Sb(3), Sc(4), and
Irr(5). The classification was verified for 34 of those galaxies for
which morphological types were found in the NASA/IPAC Extragalactic
Database (NED)\footnote{http://nedwww.ipac.caltech.edu/index.html},
and was found to be accurate to $\pm 1$ class.

\begin{figure}
\plotone{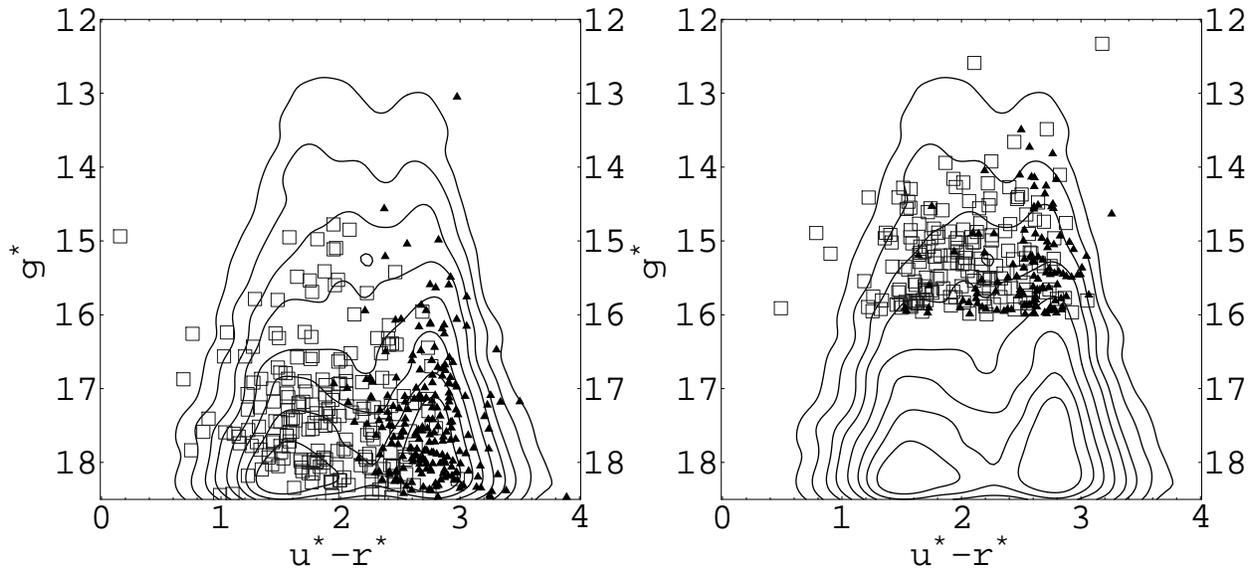}
\caption{Bimodality in the photometric galaxy sample (contours)
corresponds  to early (filled triangles) and late (open squares) types
of galaxies.  The 500 galaxies in the left panel are classified
spectroscopically,  the 287 bright galaxies on the right are
classified by visual inspection of images.\label{Cmds}}
\end{figure}

\begin{figure}
\plotone{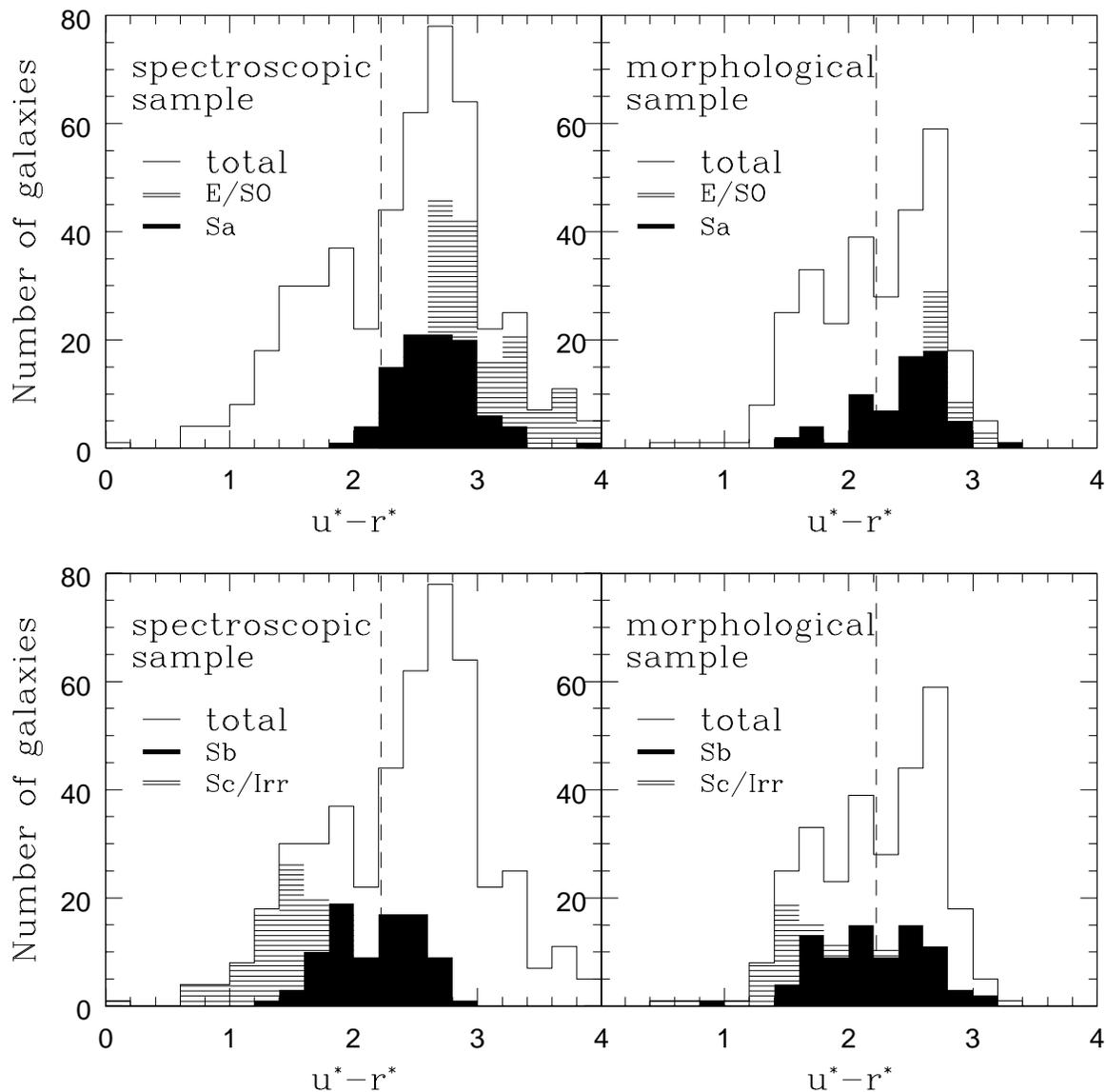}
\caption{Left: Spectroscopic classification and \urc\ color.  Right:
Morphological classification and \urc\ color. Top panels show
histograms of early type galaxies (E/SO or Sa), bottom panels for late
types (Sb or Sc/Irr).\label{ColHist}}
\end{figure}

The right panel of Figure \ref{Cmds} presents the  visually classified
galaxies  separated into early and late types as points over the
photometric sample  given in contours. \urc\ histograms for four
individual  subclasses (E/SO, Sa, Sb, and Sc/Irr) are given in the
right panel of  Figure \ref{ColHist}. Of the 117 galaxies visually
classified as E, S0 or Sa, 80\% have \urc\ colors redder than 2.22,
consistent with being early type galaxies. Of the 170 galaxies
visually classified as Sb, Sc or Irr, 112 (66\%) have colors bluer
than \urc  =  2.22 consistent with being late type galaxies. Another
16\% have  colors bluer than \urc  =  2.5, while the remaining 18\%
(30/170 galaxies)  have colors in the range 2.5 $<$ \urc $<$ 3.2 (the
reddest spiral in this sample is  NGC 4666, a dusty SABc LINER galaxy,
in the upper right corner of the right  panel of Figure \ref{Cmds},
with a B$-$V  color of 0.8 from RC2).

\subsection{   Other Parameters Sensitive to Morphology    }

The SDSS photometric pipeline calculates a number of global
morphological  parameters for every object. These include the
likelihoods of the best fit exponential  or de Vaucouleurs' profiles,
the ``texture'' parameter,  and the concentration index\footnote{A 
separate 2D bulge-disk decomposition code that allows for a S\'ersic bulge, 
exponential disk, and a bar is being developed.}. ``Texture''
measures the bilateral asymmetry of an object, but was found to  be
poorly correlated with galaxy type in its current implementation for
both  visually and spectroscopically classified galaxies and will not
be considered  further. Profile probabilities and concentration index
both correlate well with morphology. See the Appendix for more details
about the computation of these parameters.

\begin{figure}
\plotone{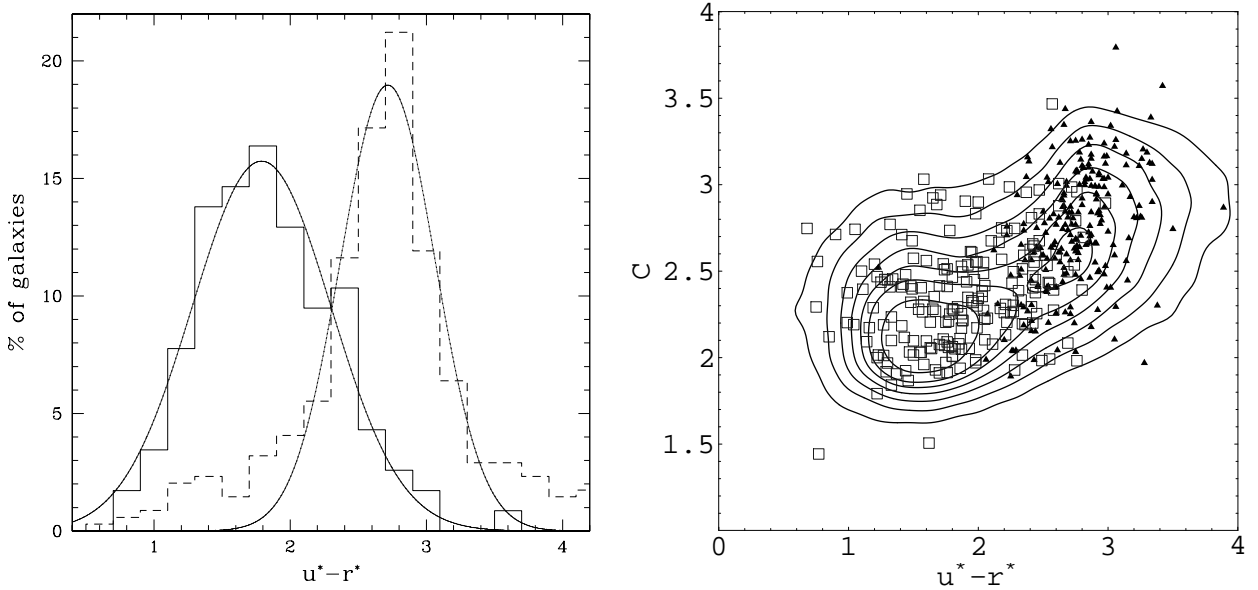}
\caption{\hspace{5mm}\urc\ color correlates with profile likelihoods
and concentration index.  Left panel: spectroscopic galaxy sample
\urc\  histograms separated into objects with $P\si{deV}>P\si{exp}$
(early type, dashed lines) and $P\si{deV}<P\si{exp}$ (late type, solid
lines) show the same bimodality as does galaxy \urc\ color. Gaussian
fits to the two histograms are given as a guide to the eye. Right
panel: concentration index vs.\ \urc. The photometric sample is given
as contours enclosing $\sigma$/4 (20.8\%) to 2$\sigma$ (95.5\%) of all
galaxies with \gM $\leqslant$ 20, in steps of $\sigma$/4 as in Figure
1. The filled triangles correspond to early spectroscopic sample
galaxies (E, S0, Sa) and the open squares to late spectroscopic sample
galaxies (Sb, Sc, Irr): early type galaxies have higher concentration
index than late types. \label{CIPP}}
\end{figure}

Using the exponential ($P\si{exp}$) and de Vaucouleurs' ($P\si{deV}$)
profile likelihoods, early type galaxies (E, S0, Sa) can be selected
by requiring $P\si{deV}>P\si{exp}$, and late types (Sb, Sc, Irr) by
$P\si{exp}>P\si{deV}$, where both likelihoods are calculated in the
$r$ band. The profile likelihood criterion is not a good
discriminator at very bright magnitudes, $\gM < 16$ (see the
Appendix). Similar conclusion was reached by Shimasaku {\em et al.},
based on a sample of 456 bright galaxies. The basic reason for the
poor performance  of profile likelihoods at the bright end is the
version of {\em Photo} pipeline used in reducing this data which bases
the profile fits on the inner profiles of galaxies, thus assigning
higher likelihoods of de Vaucouleurs fits to large, nearby spirals. In
the fainter spectroscopic sample used here this effect is
negligible. The correspondence between the profile likelihood
classification and \urc\ color is illustrated in the left panel of
Figure \ref{CIPP} based on the spectroscopic sample.  The histogram of
galaxies with  $P\si{deV}>P\si{exp}$ follows the distribution of red
galaxies (\urc $\geqslant$ 2.22),  while the histogram of galaxies
with $P\si{exp}>P\si{deV}$ follows the distribution  of blue galaxies.

The concentration index, defined as the ratio of the radii containing
90$\%$ and 50$\%$ of the Petrosian $r$ galaxy light,  $C \equiv
\case{r\si{p90}}{r\si{p50}}$, also correlates with galaxy
type. Centrally concentrated ellipticals are  expected to have larger
concentration indices than spirals. For a classical  de Vaucouleurs
profile $I\si{E}(r)  =  I\si{e} e^{-7.67((r/r_e)^{1/4}-1)}$, the
concentration  index is \about 5.5, while the exponential disks of
spirals ($I\si{S}(r)  =  I\si{s} e^{-r/r_s}$) have concentration index
\about 2.3. Both estimates correspond to the seeing--free case; the
observed values are somewhat lower. The dependence of concentration
index on morphological type found in our spectroscopic and
morphological samples is  weak, with large scatter in C for each
morphological type.  The linear correlation coefficients are 0.4 and
0.7 for the spectroscopic and morphological samples respectively, with
the probability of random samples of those sizes giving the above
correlation coefficients \about $10^{-16}$. The small correlation
coefficients and the large scatter indicate that the concentration
index is not a robust morphological separator, except  in a very crude
sense: it can be used to separate early (E, S0, Sa) from late type
(Sb, Sc, Irr) galaxies. Shimasaku {\em et al.}\ (2001) support the use
of the concentration index as morphological separator,  while noting
the large uncertainty ($\pm 1.5$ class in their 7 morphological types
system from E to Irr galaxies), and cautioning that the concentration
index cannot be used to create a pure E/S0 sample, free of Sa
contamination.

Depending on whether the completeness or reliability of a subsample of
a given type  is to be maximized, a different concentration index
separator has to be adopted.   We define the classification
\emph{reliability} as the fraction of galaxies from the selected
subsample that are correctly classified.  For example, the  \urc
$\geqslant$ 2.22 early type (E, S0, Sa) color  selection criterion
selects 343 galaxies from the spectroscopic  sample, 284 of which have
spectra consistent with an E, S0 or Sa  galaxy. Therefore the
reliability of the \urc\ color criterion for selecting spectroscopic
early types is 83\%. The \emph{completeness} is defined as  the
fraction of all galaxies of a given type from the original sample that
are selected by the classification scheme. The \urc $\geqslant$ 2.22
color selection criterion for early types selects 284 of the 290 early
type galaxies from the spectroscopic sample, resulting in a
completeness of 98\%.

Based on the brighter morphological sample, $C > 2.63$ and $C < 2.63$
give equally complete (\about 83\%) subsamples of early and late types
of galaxies, with 76\% reliability for the early and 88\% reliability
for the late type selection. Equal reliability (81\%) for  the
selection of early and late types is given by a $C  =  2.83$
separator, with completeness of 70\% and 80\% of the early and late
subsamples respectively.   Based on the spectroscopic sample, $C >
2.55$ and $C < 2.55$ give equally complete (\about 73\%) subsamples of
early and late types of galaxies, with 78\% reliability  for the early
and 66\% for the late type selection. Equal reliability (72\%) for the
early and late type  selection is given by a $C  =  2.40$ separator,
with completeness of  85\% and 54\% for the early and late subsamples
respectively. We adopt a  $C  =  2.6$ separator, which is optimized
for completeness of subsamples selected from both the spectroscopic
and morphological samples and gives equally reliable types in both.
Shimasaku {\em et al.}\ (2001) recommend a different separator for an
early-to-late cut at SO/Sa. They suggest an inverse concentration
index of 0.33 ($C \about 3$), based on a bright sample of 456
galaxies,  and optimized for low contamination. The different
definition of the  cut and its optimization make direct comparison
with our results  difficult.

The correlation of the concentration index and \urc\ color is
illustrated in the right panel of Figure \ref{CIPP}, where a $\gM <
20$ subsample is given as contours and the spectroscopic sample as
points. The limit of $\gM  =  20$ for the galaxy subsample in Figure
\ref{CIPP} was chosen to ensure low contamination and a higher
fraction of red galaxies.

\subsection{Comparison of morphology selection criteria}

The selection criteria for early and late type galaxies are compared
in  Tables 2 and 3 against the spectroscopic and visual classification
respectively.  The upper part of the tables compares selection methods
for early type galaxies,  while the lower part compares selection
methods for late type galaxies. From Table 2,  based on the
spectroscopic classification, all selection criteria have  comparable
reliability for selecting early types, but the concentration  index
selection method is only \about 68\% complete, compared to more than
95\% completeness of the profile likelihood and color selection
methods.  The color selection method is preferable in choosing
subsamples of late type galaxies, with \about 72\% completeness and
\about 96\% reliability.  The profile  likelihood method for late type
selection misses almost half of all  late type galaxies, while the
concentration index criterion is only  \about 64\% reliable.
 
Table 3 gives analogous results based on the visual classification of
the bright (\gM $<$ 16) morphological sample.  The concentration index
criterion has a comparable performance for this bright galaxy sample
as for the spectroscopic sample, with increased reliability  for
selecting early (from 68\%  to 84\%) and late (from 73\% to 85\%) type
galaxies.  The profile likelihood criterion has severe problems for
very bright  galaxies, missing 94\% of the spirals present in this
sample (see the Appendix).  For the bright morphological  sample, the
color criterion is less complete and less reliable for both early and
late type selection than the concentration index method.

\begin{figure}
\plotone{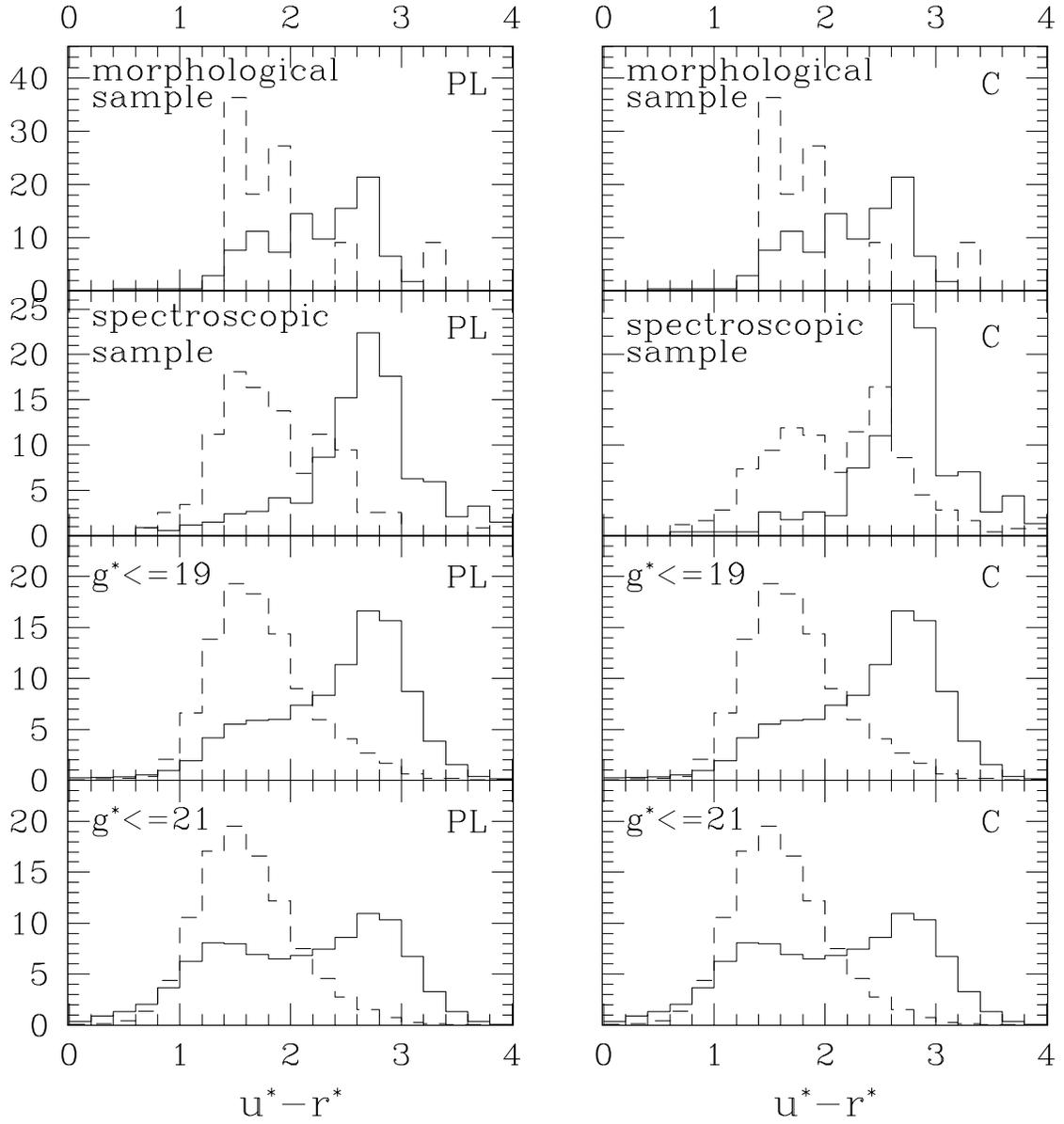}
\caption{\hspace{5mm}\urc\ color histograms of early type (solid
lines) and late type  (dashed lines) galaxies selected using the
profile likelihood (left panel) and  concentration index (right panel)
criteria.\label{CIPPchange}}
\end{figure}

Unfortunately, the color, concentration index and profile likelihoods
criteria can be compared to independent morphological classification
only at the bright end. Beyond \gM \about 18.5 (which roughly
corresponds to the spectroscopic limit of $\rM < 17.77$), we are
forced to compare the different methods between themselves.  The
correlation between color and concentration index gets weaker for
fainter magnitude samples, as illustrated in the right panel of Figure
\ref{CIPPchange}. In the spectroscopic sample, only 10\% of the
galaxies with higher concentration index characteristic of early types
($C > 2.6$) have colors bluer than $\urc = 2.22$. For $17 < \gM \leq
19$ in the photometric sample, 18\% of $C > 2.6$ galaxies have blue
colors, with the percentage increasing to 26\% and 43\% for $19 < \gM
\leq 20$ and $20 < \gM \leq 21$ respectively. When we impose a more
stringent cut for selecting early types, requiring both $C > 2.6$ and
$P\si{deV} > P\si{exp}$, for  $17 < \gM \leq 19$ we reduce the number
of selected galaxies by \about 13\%, and the fraction of blue galaxies
among those selected to 13\%. For $19 < \gM \leq 20$ and $20 < \gM
\leq 21$, 20\% and 36\% less galaxies are selected, and the blue
fraction is reduced to 17\% and 26\% respectively. Since we expect no
more than 75\% to 81\% reliability even at the bright end for early
type selection based on the concentration index criterion alone, we do
not expect the contamination to be much less than \about 20\%. It is
therefore likely that by using the concentration index criterion alone
to select early types, we are including an extra 10\% to 20\% of late
types for $19 < \gM \leq 20$ and $20 < \gM \leq 21$ respectively.

A similar trend of weakening of the correlation between profile
likelihoods and \urc\ color for faint galaxies is also observed (see
the left panel of Figure \ref{CIPPchange}).  For $\gM > 19$ there is
an increasing fraction of galaxies which are better fit by a de
Vaucouleurs' profile and have blue colors $\urc < 2.22$. For $20 < \gM
<21$, this fraction increases to 43\% of all $P\si{deV} > P\si{exp}$
galaxies, from \about 30\% for brighter magnitudes.  The profile
likelihood criterion is less discriminative for early type selection
than either the concentration index or \urc\ color criteria, at both
bright and faint magnitudes. Comparing the numbers of galaxies
selected as early types by $P\si{deV} > P\si{exp}$ alone with those
selected by both $C > 2.6$ and $P\si{deV} > P\si{exp}$, we find that
only 57\% ($17 < \gM \leq 19$), 48\% ($19 < \gM \leq 20$), and 41\%
($19 < \gM \leq 20$) are ``confirmed'' by the concentration index cut.
Thus both concentration index and color of faint galaxies seem to
suggest that the profile likelihood method for selecting early types
includes as much as extra 50\% to 60\% late type galaxies at the faint
end ($19 < \gM \leq 21$).

Since we do not expect appreciable evolution in the colors of early
and late type galaxies, and theoretical population synthesis studies
suggest that \urc\ color remains a good discriminator over the
redshift  range considered ($z \leq 0.4$), we are inclined to believe
that the profile likelihood and concentration index method, which rely
on  spatial information hard to obtain at the faint end, fail before
the color criterion does. The discrepancies between the three methods
for faint magnitudes, quantified above, support that view.

Overall, the concentration index criterion is can be used at the very
bright  end ($\gM < 16$), where galaxies are also easy to classify
both visually and spectroscopically. The profile likelihood criterion
is currently applicable  to galaxies at intermediate magnitudes ($16 <
\gM < 18$) with a somewhat low completeness for late type selection
and low reliability for early type selection. In this range the
concentration index and color criteria both give better results. The
color criterion is applicable for all magnitude ranges considered, and
we specifically recommend its use for fainter samples ($\gM > 18$).

\section{Conclusions} 

This study indicates that galaxies have a bimodal \urc\ color
distribution corresponding to early (E, S0, Sa) and late (Sb, Sc, Irr)
morphological types, that can be clearly separated by a \urc\ color
cut of 2.22, independent of magnitude. The peak-to-peak width of this
\urc\ color separation is $\gtrsim$ 1$^m.1$, about twice as large as
the sum in quadrature of the widths of each peak (where the width
includes both color errors and the intrinsic color dispersion), even
for the faintest magnitudes considered (\gM $\lesssim$ 21).  As can be
seen from Figures \ref{CmdCcd} and \ref{RiIzHist}, other color
combinations (\ri, \iz, \gr, and \ug) have much smaller peak-to-peak
separations between the two populations. The SDSS $u-r$ color is a
unique combination of an ultraviolet $u$ band bluer than the
Johnson-Morgan U band ($\lambda\si{eff}(u) = 3543$~\AA\ vs.\
$\lambda\si{eff}(U) = 3652$~\AA) and a broader and redder $r$ band
($\lambda\si{eff} = 6231$~\AA, FWHM = 1373~\AA) compared to
Johnson-Morgan V ($\lambda\si{eff} = 5505$~\AA, FWHM = 827~\AA),
allowing for a more sensitive comparison of the blue and red portions
of the galaxy spectra, relevant for isolating star formation rates.
The \uM\ and \rM\ filters always bracket the Balmer break, and the
theoretical tracks of galaxy color evolution with redshift in the \ug\
vs.\ \gr\ color--color diagram are parallel to the \urc = constant
lines up to z \about\ 0.4, supporting the claim that this separation
is applicable to samples at all redshifts relevant for the main sample
of SDSS galaxies. We find evidence for the evolution of  the \urc\
color distribution with fainter magnitudes, indicative of the presence
of larger fractions of bluer galaxies at redshifts of z \about 0.4.

Among the SDSS parameters calculated automatically by the photometric
pipeline \emph{Photo}, concentration indices and profile likelihoods
are also sensitive to morphology and correlate  with \urc\ color.
Unlike the concentration index and profile likelihood methods,  the
\urc\ color separation does not require well resolved images.

Since star formation rate (SFR) is one of the defining characteristics
of  the Hubble sequence, it is expected that \urc\ color, as an
indicator of  recent star formation, correlates with Hubble
type. Other defining parameters for the Hubble types are the
bulge-to-disk (B/D) ratio, and the tightness of the spiral
pattern. Those three parameters can be effectively reduced to  two
underlying ones --- the star formation history and the mass
distribution  of the galaxy. The B/D ratio, the concentration index
and the profile likelihoods  are just three ways of measuring the mass
distribution using the stellar light, while the tightness of the
spiral pattern is presumably dependent on the mass distribution  as
well as recent star formation. Moreover the mass distribution and star
formation are probably not strictly independent; for example, we found
correlations between the color as a measure of star formation, and the
concentration index and the profile likelihoods as measures of the
mass distribution. In view of this dependence of Hubble types on  two
underlying internal (as opposed to environmental) parameters, it is
not surprising to see that neither of the measures we use gives  a
perfect correspondence to morphology. The \urc\ color, concentration
index and profile likelihoods, as independent, quantitative indicators
of morphology are more accessible and less subjective than Hubble
types,  and thus more beneficial to the studies of galaxy properties
and formation.

\vskip 0.5in \leftline{Acknowledgments}

Iskra Strateva and Gillian Knapp are grateful to NASA for support  via
NAG5-3364.  Michael Strauss acknowledges the support of NSF grant
AST00-71091. The authors wish to thank the referee, Michael Fioc, for
his  insightful comments and suggestions.

The Sloan Digital Sky Survey (SDSS) is a joint project of The
University of Chicago, Fermilab, the Institute for Advanced Study, the
Japan Participation Group, The Johns Hopkins University, the
Max-Planck-Institute for Astronomy (MPIA), the Max-Planck-Institute
for Astrophysics (MPA), New Mexico State University, Princeton
University, the United States Naval Observatory, and the University of
Washington. Apache Point Observatory, site of the SDSS telescopes, is
operated by the Astrophysical Research Consortium (ARC).  Funding for
the project has been provided by the Alfred P. Sloan Foundation, the
SDSS member institutions, the National Aeronautics and Space
Administration, the National Science Foundation, the U.S.\ Department
of Energy, the Japanese Monbukagakusho, and the Max Planck
Society. The SDSS Web site is http://www.sdss.org/.

This research has made use of the NASA/IPAC Extragalactic Database
(NED) which is  operated by the Jet Propulsion Laboratory, California
Institute of Technology,  under contract with the National Aeronautics
and Space Administration.


\appendix

\section{Model Quantities}

The optimal measure of the total flux associated with an object is the
result of the convolution of the image with a matched filter.  Even if
the matched filter used is not an accurate representation of the shape
of the object, this gives an unbiased measure of the color of the
object if the same matched filter is used in each band.  With this in
mind, the SDSS \emph{Photo} pipeline performs three model fits to each
object in every band: a Point Spread Function (PSF), a pure de
Vaucouleurs profile, and an exponential disk; the galaxy models are
convolved with the local PSF.  In each case, the fit is done to the
two-dimensional data, and the galaxy models allow for an arbitrary
scalelength, axial ratio, and position angle.  These fits are carried
out by minimizing $\chi^2$ not over each pixel (which would be
terribly time-consuming), but over a series of {\it sectors}, which
divide the image into radial and angular bins (Lupton et al.~2001).
An error is associated with the flux in each sector, based on photon
statistics and the measured pixel variance within the sector.  Galaxy
colors are measured by applying the best-fit model of an object in the
$r$ band to the other bands, thus measuring the flux through the same
effective aperture.

  Of course, real galaxies do not necessarily follow pure exponential
or de Vaucouleurs profiles: they have composite profiles, spiral arms,
and other features not included in these models.  Thus the likelihoods
of the model fits tend to be low for well-resolved, high
signal-to-noise ratio galaxies, and the likelihood ratio of
exponential and de Vaucouleurs fits is a poor measure of morphology.
Moreover, the version of \emph{Photo} used in these reductions, weighs
strongly the inner parts of galaxies while performing the profile
fits, which overestimates the contribution of the de Vaucouleurs
bulge. Work is on-going to include model uncertainties in the error
associated with the photometry in each sector in \emph{Photo}, which
should make these fits more meaningful at the bright end.

\begin{figure}
\plotone{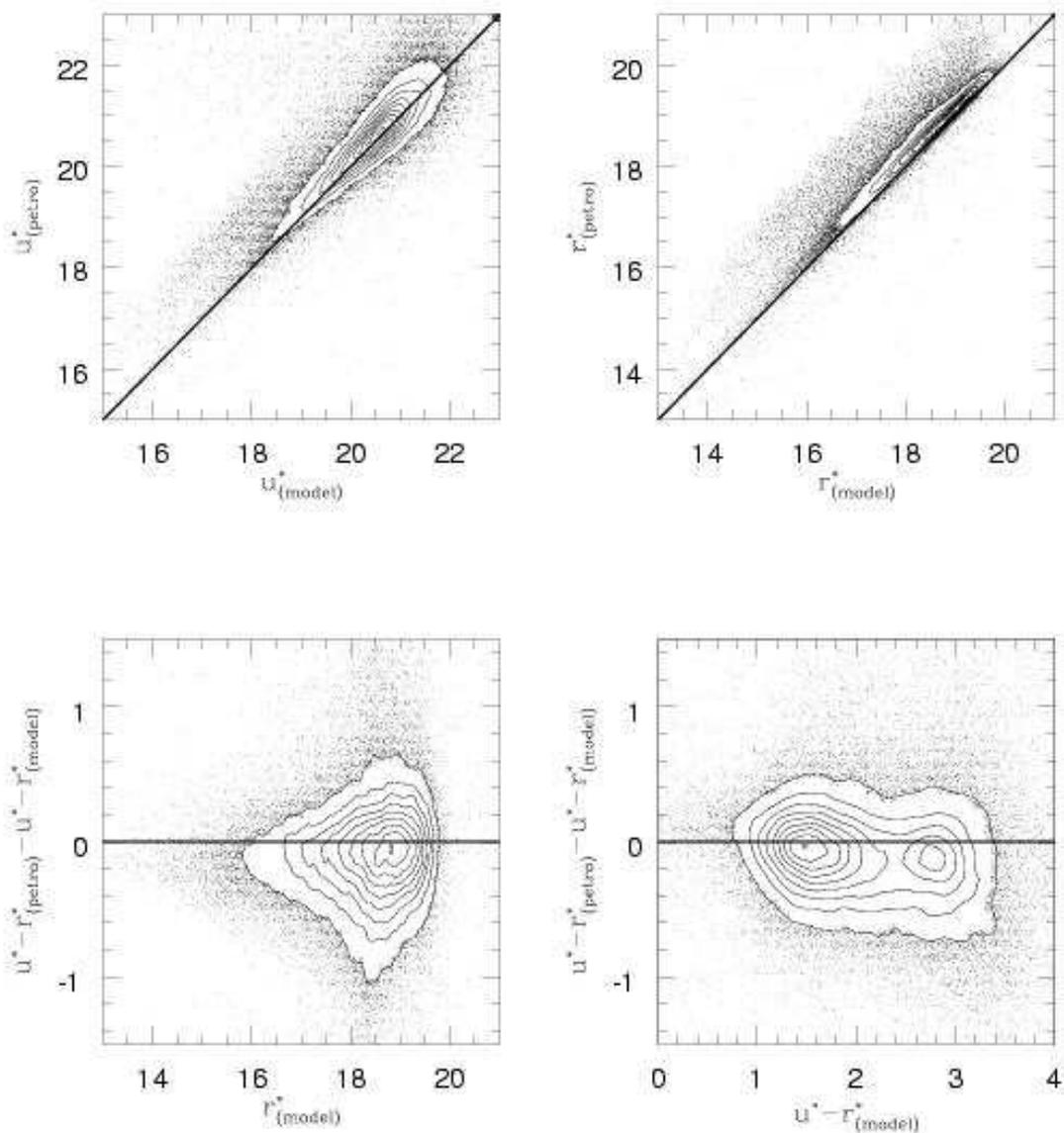}
\caption{Model and Petrosian magnitudes. The contour curves are
linearly  spaced isodensity curves, the outliers are shown as points.
Top left panel: $\uM\si{model}$ vs.\ $\uM\si{petro}$. Top right panel:
$\rM\si{model}$ vs.\ $\rM\si{petro}$. Lower panels: the difference
between  Petrosian and model \urc\ color as a function of
$\rM\si{model}$ magnitude  (left) and $(\urc)\si{model}$ color
(right). For more details see the  Appendix.\label{ModPetro}}
\end{figure}

The model flux is calculated by integrating all the light under the
best fit profile and consequently is not equal to the Petrosian flux
for a given galaxy.  A comparison between model and Petrosian
magnitudes in the \uM\ and \rM\ bands  and \urc\ color is given in
Figure \ref{ModPetro}. The majority of galaxies are enclosed  in
linearly spaced, isodensity contours, with the outliers shown as
points.  At the faint end ($\uM > 19$ and $\rM >16$), the model and
Petrosian magnitudes in \uM\  (left panel) and \rM\ (right panel) are
almost identical, with the majority of  galaxies lying close to the
model equals Petrosian magnitude line. At the bright  end, however, a
large fraction of the galaxies have fainter Petrosian magnitudes by as
much as $1^m$. Those are galaxies better fit by the de Vaucouleurs
profile for which the Petrosian flux is not equal to the total flux.
The lower panel of Figure \ref{ModPetro} show the difference in \urc\
color measured using the model and Petrosian magnitudes as a function
of model magnitude and model \urc\ color.  $\urc\si{petro}$ is \about
$0^m.15$ bluer than $\urc\si{model}$, with the difference being  more
pronounced for ellipticals than spirals (lower right panel).  As
pointed out by Yasuda et al.~(2001), the model fits are heavily
weighted towards the bright cores of galaxies, and thus de-emphasize
the bluer disk component of ordinary spiral galaxies.  This is
consistent with somewhat redder model than Petrosian colors.

\begin{deluxetable}{cccccccc} 
\tablewidth{0pc}
\tablenum{1}
\tablecaption{Photometric error coefficients and sample errors} \label{tab1}
\tablehead{\colhead{Band} &\colhead{$c_1$} &\colhead{$c_2$} 
&\colhead{m  =  16} &\colhead{m  =  18} &\colhead{m  =  20} &\colhead{m  =  21} &\colhead{m  =  22} } 
\startdata 
\uM & $0.021$   & $1.3 \times 10^{-5}$  & 0.04 & 0.07 & 0.15 & 0.23 & 0.35 \\
\gM & $0.002$   & $3.6 \times 10^{-6}$  & 0.01 & 0.02 & 0.04 & 0.06 & 0.09 \\
\rM & $0.008$   & $5.3 \times 10^{-6}$  & 0.02 & 0.03 & 0.06 & 0.09 & 0.14 \\
\iM & $0.002$   & $7.0 \times 10^{-6}$  & 0.01 & 0.03 & 0.07 & 0.11 & 0.18 \\
\zM & $0.006$   & $2.0 \times 10^{-5}$  & 0.04 & 0.09 & 0.20 & 0.32 & 0.51 \\
\enddata
\tablecomments{\hbox{$\Delta m \approx c_1+c_210^{0.2m}$} fits for $c_2$ were obtained
using \about 5000 randomly selected galaxies from the photometric sample, with
$c_1$ set to the median error for $m < 16$.}
\end{deluxetable}

\begin{deluxetable}{ccc} 
\tablewidth{0pc}
\tablenum{2}
\tablecaption{Comparison between classification methods for the spectroscopic sample} \label{tab2}
\tablehead{\colhead{Selection Rule} &\colhead{Completeness} &\colhead{Reliability}} 
\startdata 
\urc $\geqslant$ 2.22  & $98\%$   & $83\%$  \\
$P\si{deV}>P\si{exp}$  & $96\%$   & $76\%$  \\
$C>2.6$                & $68\%$   & $81\%$  \\
\hline
\urc $<$ 2.22          & $72\%$   & $96\%$  \\
$P\si{exp}>P\si{deV}$  & $55\%$   & $90\%$  \\
$C<2.6$                & $77\%$   & $64\%$  \\
\enddata
\tablecomments{The exponential and de Voucouleurs law likelihoods and the
radii used in computing the concentration index are all measured in the \rM\ band.}
\end{deluxetable}

\begin{deluxetable}{ccc} 
\tablewidth{0pc}
\tablenum{3}
\tablecaption{Comparison between classification methods for the morphological sample} \label{tab3}
\tablehead{\colhead{Selection Rule} &\colhead{Completeness} &\colhead{Reliability}} 
\startdata 
\urc $\geqslant$ 2.22  & $80\%$   & $62\%$  \\
$P\si{deV}>P\si{exp}$  & $99\%$   & $42\%$  \\
$C>2.6$                & $84\%$   & $75\%$  \\
\hline
\urc $<$ 2.22          & $66\%$   & $83\%$  \\
$P\si{exp}>P\si{deV}$  & $6\%$    & $91\%$  \\
$C<2.6$                & $81\%$   & $88\%$  \\
\enddata
\end{deluxetable}

\newpage
\newpage
\newpage 
\newpage
\newpage
\newpage
\newpage
\newpage
\newpage
\newpage


\begin{thebibliography}{}
\bibitem[Blanton 2001]{Blanton01}Blanton, M., Dalcanton, J., Eisenstein, D., {\em et al.}\ 2001, \aj, 121, 2358B
\bibitem[Bernardi 2001]{Ber01}Bernardi M., Sheth, R. K., Annis, J.,  {\em et al.}\ 2001, in preparation
\bibitem[Brown {\em et al.}\ 2000]{Brown00}Brown, M. J. I., Webster, R. L., Boyle, B. J. 2000, \mnras, 317, 782
\bibitem[Buta \& Williams 1994]{Buta94}Buta, R., \& Williams, K. L., 1994, \aj, 109, 543B
\bibitem[Cheeseman \& Stutz(1996)]{bayes1} Cheeseman, P., Stutz, J., ``Bayesian Classification (AutoClass): 
Theory and Results'', in {\it Advances in Knowledge Discovery and Data Mining}, Usama M. Fayyad, Gregory 
Piatetsky-Shapiro, Padhraic Smyth, \& Ramasamy Uthurusamy, Eds. AAAI Press/MIT Press, 1996
\bibitem[Connaly {\em et al.}\ 2001]{co01}Connaly A., Johnston, D., Dodelson, S., {\em et al.}\ 2001, \apj, submitted
\bibitem[de Vaucouleurs 1961]{deV61}de Vaucouleurs, G. 1961, \apjs, 5, 233
\bibitem[Dodelson {\em et al.}\ 2001]{do01}Dodelson, S., Narayanan, V., Tegmark, M., {\em et al.}\ 2001, \apj, submitted
\bibitem[Eisenstein {\em et al.}\ 2001]{Eis01}Eisenstein, D., Annis, J., Gunn, J. E., {\em et al.}\ 2001, \aj, submitted
\bibitem[Fan 1999]{Fan99}Fan, X. 1999, AJ, 117, 2528
\bibitem[Fan {\em et al.}\ 1999]{F99}Fan, X., Strauss, M., Schneider, D. {\em et al.}\ 1999, AJ, 118, 1
\bibitem[Ferreras {\em et al.}\ 2000]{Fer99}Ferreras, I., Cayon, L., Martinez-Gonalez, E. {\em et al.}\ 1999, \mnras, 304, 319
\bibitem[Fioc {\em et al.}\ 1999]{Fioc99}Fioc, M., \& Rocca-Volmerange, B. 1999, \aap, 351, 869
\bibitem[Fioc {\em et al.}\ 1997]{Fioc97}Fioc, M., \& Rocca-Volmerange, B. 1997, \aap, 326, 950F
\bibitem[Finlator {\em et al.}\ 2000]{Finl00}Finlator, K., Ivezi\'{c}, \v{Z}., Fan X. {\em et al.}\ 2000, \aj, 120,2615
\bibitem[Fischer {\em et al.}\ 2000]{Fisch00}Fischer, P., McKay, T., Sheldon, E., {\em et al.}\ 2000, \aj, 120, 1198
\bibitem[Fukugita {\em et al.}\ 1995]{Fuk95}Fukugita, M., Shimasaku, K., \& Ichikawa, T. 1995, \pasp, 107, 945
\bibitem[Fukugita {\em et al.}\ 1996]{F96}Fukugita, M., Ichikawa, T., Gunn, J. E., {\em et al.}\ 1996, AJ, 111, 1748
\bibitem[Goebel {\em et al.}\ 1989]{bayes0}Goebel, J., {\em et al.}\ 1989, A\&A, 222, L5 
\bibitem[Gunn {\em et al.}\ 1998]{camera}Gunn , J. E., Carr, M., Rockosi, C. {\em et al.}\ 1998, \aj, 116, 3040
\bibitem[Hanson, Stutz \& Cheeseman(1991)]{bayes2}
Hanson, R., Stutz, J., Cheeseman, P. 1999, ``Bayesian Classification Theory'', 
Technical Report FIA-90-12-7-01,  NASA Ames Research Center, Artificial 
Intelligence Branch
\bibitem[Hubble 1936]{Hub36}Hubble, E. 1936, {\it The Realm of the Nebulae}, Oxford University Press
\bibitem[Humason, M. L. 1936]{Hum36}Humason, M. L. 1936, \apj, 83, 10
\bibitem[Ivezi\'{c} \& Elitzur 2000]{bayes4}Ivezi\'{c}, \v{Z}. \& Elitzur, M. 2000, ApJ, 534, L93
\bibitem[Kauffmann{\em et al.}\ 1996]{Kauf96} Kauffmann, G., Charlot, S., \& White, S. 1996, \mnras, 283, L117
\bibitem[Kennicutt 1992a]{k1} Kennicutt, R. C., 1992a, \apjs, 79, 255
\bibitem[Kennicutt 1992b]{k2} Kennicutt, R. C., 1992b, \apj, 388, 310
\bibitem[Kochanek 2000]{koch} Kochaneck, C. S., Pahre, M. A., \& Falco, E. E. 2000, astro-ph/001458
\bibitem[Krisciunas, Margon \& Szkody 1998]{KMS98} Krisciunas, K., Margon, B., \&  Szkody P. 1998, \pasp, 110, 1342
\bibitem[Lupton, Gunn \& Szalay 1999]{LGS99} Lupton, R. H., Gunn, J. E., \& Szalay, A. 1999,  \aj, 118, 1406
\bibitem[Lupton {\em et al.}\ 2000]{Lupton00} Lupton, R. H., Ivezi\'{c}, \v{Z}., Knapp, G. R., {\em et al.}\ 2001, in preparation
\bibitem[Morgan \& Mayall 1957]{MM57} Morgan, W. W. \& Mayall, N. U. 1957, \pasp, 69, 409
\bibitem[Oke \& Gunn 1983]{OG83}Oke, J. B., \& Gunn, J. E. 1983, \apj, 266, 713
\bibitem[Schlegel, Finkbeiner \& Davis 1998]{SFD98} Schlegel, D., Finkbeiner, D. P. \& Davis, M. 1998, ApJ 500, 525
\bibitem[Shimasaku {\em et al.}\ 2001]{Shi00}Shimasaku, K., Fukugita, M., Doi, M., {\em et al.}\ 2001, \aj, submitted 
\bibitem[Strauss {\em et al.}\ 2001]{Str01}Strauss, M., Weinberg, D., Lupton, R. H.,  {\em et al.}\ 2001, in preparation
\bibitem[Stoughton {\em et al.}\ 2001]{EDR}Stoughton, C., Adelman, J. K., Blanton, M., {\em et al.}\ 2001, in preparation
\bibitem[Szalay {\em et al.}\ 2001]{sz01}Szalay, A., Jain B., Matsubara, T., {\em et al.}\ 2001, in preparation
\bibitem[Tegmark {\em et al.}\ 2001]{tg01}Tegmark, M., Dodelson, S., Eisenstein, D., {\em et al.}\ 2001, \apj, submitted
\bibitem[Yasuda {\em et al.}\ 2001]{Yas00}Yasuda, N., Fukugita, M., Narayanan, V., {\em et al.}\ 2001, \aj, in press
\bibitem[York {\em et al.}\ 2000]{York} York, D. G., Adelman, J., Anderson, J. G., {\em et al.}\ 2000, \aj, 120, 1579
\bibitem[Zehavi {\em et al.}\ 2001]{zh01}Zehavi, I., Blanton, M., Frieman, J. A., {\em et al.}\ 2001, \apj, submitted
\end{thebibliography}
\end{document}